\documentclass[10pt,twocolumn,amsmath,amssymb,aps,physrev,nofootinbib]{revtex4-2}
\usepackage{amsmath}
\usepackage{multirow}
\usepackage{graphicx}
\usepackage{dcolumn}
\usepackage{bm}

\begin{document}

\title{Deriving Tsallis Entropy from First Principles} 
\author{J. A. S. Lima$^{1}$} \email{jas.lima@iag.usp.br}
 \author{M. H. Benetti$^{1}$} \email{mhbenetti@usp.br}

\affiliation{$^{1}$Departamento de Astronomia (IAGUSP), Universidade de S\~ao Paulo, 05508-090 S\~ao Paulo SP,  Brasil}

\date{\today}

\begin{abstract}

The kinetic approach for q-distribution  is extended through a {\it ``bottom-up”} procedure to the corresponding entropic formula. In this way,  the validity of Tsallis' formula, $S_q(W)=k_B\ln_q(W)$, is established from first principles. The resulting $q$-entropy is  additive and the zeroth law of thermodynamics is automatically satisfied because the conventional identification $\beta = (k_BT)^{-1}$ remains valid. Entropy, energy, and all relevant quantities for both discrete and continuous cases, are evaluated using standard (linear) averages. The {\it ``bottom-up”} and {\it ``top-down"} approaches yield the same $q$-distribution for velocities.  The standard results are naturally recovered in the limit $q \rightarrow 1$.

\end{abstract}

\maketitle

\section{Introduction}
The structure of the kinetic theory of gases (\textbf{KTG}) rests on two pillars: (i) Maxwell's seminal derivation of the equilibrium velocity distribution function $F(v_x,v_y,v_z)$, and (ii) Boltzmann combinatorial formulation of entropy. Initially, Maxwell \cite{Maxwell1860} considered a homogeneous and isotropic gas consisting of $N$ rigid, spherical particles confined within a volume $V$, undergoing elastic collisions, and supposed in equilibrium at temperature $T$. The calculations were based on two essential hypotheses: 
\begin{description}
  \item[M1 (Independence of velocity components)]
  \begin{equation}\label{E1}
    F(\mathbf v)=f(v_x)\,f(v_y)\,f(v_z).
    \end{equation}
  \item[M2 (Isotropy of velocity space)]
  \begin{equation}\label{E2}
    F(\mathbf v)=F\!\left(\sqrt{v_x^2+v_y^2+v_z^2}\right)\equiv F(v).
    \end{equation}
\end{description}
Combining (\ref{E1}) and (\ref{E2}), the functional differential equation guided Maxwell to the Gaussian velocity  distribution: 
\begin{equation} \label{E3}
F(v) = n\left(\frac{m}{2\pi k_B T}\right)^{\!3/2}
  \exp\!\left(-\,\frac{m\,v^2}{2k_B T}\right),
\end{equation}
a result laying the foundations for the probabilistic approach to kinetic theory, thereby paving the way for Boltzmann's formulation of statistical mechanics. Probably inspired by Maxwell distribution and his work on collisions, Boltzmann \cite{Boltzmann1877} introduced the celebrated relation $S \propto \ln W$, where $W$ represents the total number of microscopic configurations compatible with the macroscopic state of the system. In the case of an ideal gas, $W$ is a function of the energy, number of particles, and volume, namely $W \equiv W(E, N, V)$.

Nevertheless, it was Planck who provided the first general derivation of the Boltzmann entropy formula and demonstrated how the Boltzmann constant arises from the integration procedure. As he emphasized: ``This universal integration constant is the same for a terrestrial as for a cosmic system” (see for instance \cite{Planck1901,Planck1914}). 

 Planck’s reasoning formed a methodological and pedagogical bridge from Maxwell velocity distribution  to Boltzmann’s formula, based on two very simple ideas.  Like Boltzmann, he also assumed that the entropy $S(W)$ is a universal function of its argument, where $W$ represents the probability associated with the total number of microstates. For two independent systems, the Maxwellian (Gaussian) condition must apply. Furthermore, guided by Boltzmann’s logarithmic suggestion, Planck assumed that the total entropy of a composite system is the sum of the entropies of its components. Thus, in the spirit of Maxwell approach, his derivation also rested on two fundamental postulates:
\begin{description}
  \item[P1 (Independence of microstates)]
  \begin{equation}\label{E4}
    W=W_{12}=W_1\,W_2\,,
  \end{equation}
  \end{description}
\begin{description}
  \item[\textbf{P2 (Additive entropy)}] 
  \begin{equation}\label{E5}
    S(W)=S(W_{12})=S_1 (W_1) + S_2(W_2)\,.
    \end{equation}
\end{description}
 
Based on both postulates, Planck derived  a  functional equation whose solution is - up to the universal multiplicative constant -  just the entropy formula $S(W)$  proposed by Boltzmann:
\begin{equation} \label{E6}
S=k_B\ln W\,.    
\end{equation}
More interestingly, the above formula was derived with no specification to the system of interest. In particular, it was  obtained without discussing  the combinatorial Boltzmann probability approach, originally applied for a classical gas of N particles. 

 Note also that the route from Maxwell to Planck  share a common structural assumption, namely:  \emph{statistical independence} (\textbf{M1} and \textbf{P1}). This kind of  approach provided  the link uniting the probabilistic insight introduced by Maxwell to the Boltzmann-Planck's statistical mechanics formulation.  This sort of continuity underlies what is being referred to as a \textit{“bottom-up”} research program, that is, from \textbf{KTG} to Boltzmann Statistical Mechanics (see for instance \cite{LD2020}). In principle, the continuity of this \textit{“bottom-up”} evolutive research program providing the foundations of the systematic development from \textbf{KTG} to statistical mechanics may also be useful in different frameworks. 

At this point, since velocity power-laws distributions are being highly considered in the current literature -- see \cite{Link} for an updated bibliography -- it is natural to ask  whether the lessons related to the  development from Maxwell to Boltzmann-Planck complementary approaches may also be useful nowadays. This occurs due to the entropic formula proposed by Tsallis 

\begin{equation}
S_q(W)=k_B\ln_q W\,, 
\end{equation} 
as an extension of the Boltzmann expression, where the $q$-parameter quantifies the departure from additivity \cite{Tsallis1988,Tsallis1998}.  

In this context, the central aim of our letter is to demonstrate that the Neo-Maxwellian approach  to power-laws \cite{SPL1998,LSP2001,LB2025a,LB2025b}, can also be explored one step further, by deducing the Tsallis entropic formula based on a Neo-Planckian approach as defined here. Hence, in the same spirit that the Maxwell postulate \textbf{M1} has been modified to obtain power velocity distributions, we revisit the  pair of postulates by Planck, in a generalized $q$-deformed setting. As we shall see, beyond completing the “bottom-up” program of Tsallis’ statistics, the connection with thermodynamics (particularly the zeroth law) is also clarified. In particular, the basic criticism related to the zeroth law of thermodynamics in the Tsallis framework is fully eliminated.

\section{Power-Laws Velocity Distributions}

To begin with, let us summarize the first step where the so-called Neo-Maxwellian approach has been applied to derive more general equilibrium distributions (power-laws). In a series of articles \cite{SPL1998,LSP2001,LB2025a,LB2025b}  we have shown that replacing Maxwell’s assumption for the components of velocity distributions (\textbf{M1}) by a $q$-deformed composition rule, but preserving isotropy (\textbf{M2}), two classes of power-laws are  derived. For an ideal gas of $N$ particles at temperature $T$ in a closed box, the first Neo-Maxwellian postulate can be written as:  

\begin{description}
  \item[\textbf{M1* (Correlation of velocity components)}]
  \begin{equation}\label{E7}
    F(\textbf{v})= e_q\left[\ln_q f(v_x)+\ln_q f(v_y)+\ln_q f(v_z)\right],
     \end{equation}
\end{description}
where the $q$-expressions above are defined by:
\begin{equation}\label{E8}
  e_q(x)=\big[\,1+(1-q)x\,\big]^{\frac{1}{1-q}},
  \qquad
  \ln_q(x)=\frac{x^{\,1-q}-1}{1-q}.
\end{equation}
These q-functions  play the role of generalized Euler exponential and  logarithmic functions, in virtue of the limiting properties:  $\lim_{q\to1}e_{q}(x)=\exp(x)$, and $\lim_{q\to1}\ln_{q}(x)=\ln x$. The identity, $e_{q}[\ln_{q}(x)]=\ln_{q}[e_{q}(x)]\equiv x$,  implies that both classes are also inverse functions. In addition, expression (\ref{E7}) recovers the Maxwell postulate in (\ref{E1}) when $q$ goes to unity. Notice that the second postulate assuming isotropy of velocity space is the same of Maxwell (\textbf{M2*= M2}). Under these assumptions, the 3D Tsallis velocity distribution takes the form: 

\small
\begin{align}\label{E9}
  F(v)
  &= n\left(\frac{m(q-1)}{2\pi k_B T}\right)^{\!3/2}
     \frac{\,\Gamma\!\left(\frac{1}{q-1}+\frac{5}{2}\right)}{\Gamma\!\left(\frac{1}{q-1}+1\right)}
     \left[\,1+(1-q)\frac{m v^2}{2k_B T}\,\right]^{-\frac{1}{1-q}},\nonumber
       \\[6pt]
  F(v)
  &= n\left(\frac{m(1-q)}{2\pi k_B T}\right)^{\!3/2}
     \frac{\,\Gamma\!\left(\frac{1}{1-q}\right)}{\Gamma\!\left(\frac{1}{1-q}-\frac{3}{2}\right)}
     \left[\,1+(1-q)\frac{m v^2}{2k_B T}\,\right]^{-\frac{1}{1-q}},
    \end{align}
\normalsize
for the short-tail ($q>1$) and fat-tail ($q<1$) distributions, respectively, both expressions reduce to the celebrated Maxwellian kinetic distribution when $q\to1$.

\section{Basic Postulates and Tsallis Entropy}

Motivated by the Neo-Maxwellian construction in velocity space and also for methodological consistency, we now  \emph{replace} the standard factorization Planck's postulate [see \textbf{P1} in (4)] by the $q$-factorization for the microstates as suggested by (\ref{E7}). So, our new postulate becomes:
\begin{description}
  \item[P1* (Correlation of microstates)]
\begin{equation}\label{E10}
W=e_q\!\big(\ln_q W_1 \;+\; \ln_q W_2\big).
\end{equation}
\end{description}
As expected, we see that $W = W_1W_2$ in the limit $q\rightarrow 1$. However, since the second Maxwell's postulate (isotropy) was maintained, we also preserve the entropic additivity. Therefore, the second Neo-Planckian postulate reads:

\begin{description}
  \item[P2* (Additive entropy)]
\begin{equation}\label{E11}
S_q(W)= S_q(W_1) \;+\; S_q(W_2).
\end{equation}
\end{description}
These postulates imply that the functional equation to be solved now takes the form:
\begin{equation}\label{E12}
   S_q(W) = S_q\!\Big[e_q\!\big(\ln_q W_1 \;+\; \ln_q W_2\big)\Big]
   \;=\; S_q(W_1) + S_q(W_2).
\end{equation}
Differentiating \textit{w.r.t.} $W_1$ (keeping $W_2$ fixed) we obtain
\begin{equation}\label{E13}
S_q'(W)\,W^{q}W_1^{-q}=S_q'(W_1)\,\,\,\leftrightarrow\,\, S_q'(W)W^{q}=S_q'(W_1)W_1^{q},
\end{equation}
where all primes denote ordinary derivatives. Similarly, differentiating (\ref{E10}) \textit{w.r.t.} $W_2$ (keeping $W_1$ fixed) yields:
\begin{equation}\label{E14}
S_q'(W)\,W^{q}W_2^{-q}=S_q'(W_2)\,\,\,\leftrightarrow \,\,\,S_q'(W)W^{q}=S_q'(W_2)W_2^{q}.
\end{equation}
Thus, differential constraints (\ref{E13})-(\ref{E14}) imply that the functional $S_q(W)$ satisfies the product:
\begin{equation}\label{E15}
S_q'(W)W^{q}=S_q'(W_1)W_1^{q}=S_q'(W_2)W_2^{q}     
\end{equation}
and, therefore, $S'(W)W^{q}=K$, where $K$ must be a constant, or equivalently,
\begin{equation}\label{E16}
 \frac{dS_q}{dW} = KW^{-q} \leftrightarrow S_q(W)= K\frac{W^{1-q}}{1-q} + C.
\end{equation}
By taking $S_q(W=1) = 0$ yields $C = -\frac{K}{1-q}$. Finally, by inserting this value in the above equation we obtain:
\begin{equation}\label{E17}
S_q(W)={K} \frac{W^{\,1-q} - 1}{1- q} = K\ln_q W.
\end{equation}
Like Planck, we also choose the constant physical scale, as being the Boltzmann constant,  $K=k_{B}$, since in the limit $q=1$,  $S_1=k_B\ln W$. Hence, the ``bottom-up"\, approach developed here provides a simple first principles Tsallis' entropic formula. 

\begin{equation}\label{E18}
S_q(W)=k_B\,\ln_q W, 
\end{equation}
together an additional extra bonus, namely: for two systems in thermal contact
\begin{equation}\label{E19}
S_q(1+2) = S_q(1) + S_q(2).
\end{equation}
This additive microcanonical result for Tsallis entropy  has been obtained through  an extension of the first Planck postulate (\textbf{P1*}), in line with our earlier derivation of a power-law velocity distribution based on \textbf{M1*} \cite{SPL1998,LSP2001,LB2025a}. It aligns remarkably well with the standard \textit{“bottom-up”} line of development, initiated by Maxwell and subsequently advanced by Boltzmann and Planck.  

Notice that in the Tsallis approach, instead of the above additive property, the hypothesis of statistical independence of the microstates adopted in the literature gives rise to the non-additive result:
\begin{equation}\label{E20}
S_q(1\!+\!2)=S_q(1)+S_q(2)+\frac{1-q}{k_B}\,S_q(1)\,S_q(2).
\end{equation}

The above non-additivity is a pure consequence for the original Maxwell-Boltzmann–Planck independence condition, plus an entropic \textit{``top-down"} postulate for $S_q$. In other words, the non-additivity arises from a \textit{pure algebraic condition}, closely  related to the  independence of the microstate probabilities ($W=W_1W_2$). 

Nevertheless, it is well known that power-law distributions representing deviations from Gaussianity are inherently correlated, and, as such, cannot arise from statistical independence of velocity components. In principle, from a more profound point of view, some conceptual difficulties would be expected  when a \textit{“top-down”} approach is assumed. Naturally, this new perspective also bears on the issue of $S_q (W)$ and its thermodynamic connection.

\section{Solving the zeroth law problem of the $S_q$-Entropy}

To our knowledge, Nauenberg’s was the first to  seriously criticize the $q$-entropy  based on its thermodynamic connection \cite{Nauenberg2003,Tsallis2004,Nauenberg2004}. He argued that by keeping  the statistical independence of microstates, $W=W_1W_2$, the entropic Tsallis' formula leads directly to the non-additivity in ($\ref{E20}$), which is not compatible with the zeroth law of thermodynamics.  In this case, for two interacting systems that can exchange energy but are otherwise isolated:

\begin{equation}\label{E21}
E=E_1+E_2=\text{const}.
\end{equation}
An entropy maximum at fixed $E$ requires $\delta S_q=0$ under $\delta E_1+\delta E_2=0$. Thus, from Eq. (\ref{E20}) we have 
\begin{equation}\label{E22}
\left(\frac{\partial S_q(1)}{\partial E_1}\right)
\!\!\Bigg[1+\frac{1-q}{k_B}\,S_q(2)\Bigg]
=
\left(\frac{\partial S_q(2)}{\partial E_2}\right)
\!\!\Bigg[1+\frac{1-q}{k_B}\,S_q(1)\Bigg].
\end{equation}
Note that $S_q=S_q(E,V,N)$, and, as such, all
partial derivatives with respect to $E$ are taken at fixed $V$ and $N$ for each
subsystem. Now, defining the standard physical  temperature 
\begin{equation}\label{E23}
\frac{1}{T}\;\equiv\;\frac{1}{k_B}\,\left(\frac{\partial S_q}{\partial E}\right)_{V,N},
\end{equation}
and combining with Eq. (\ref{E21}), Nauenberg obtained the relation
\begin{equation}\label{E24}
\left[1+(1-q)\,\frac{S_q(1)}{k_B}\right]\,T_1
=
\left[1+(1-q)\,\frac{S_q(2)}{k_B}\right]\,T_2.
\end{equation}
Then, for two systems in thermal contact, the temperatures $T_1$ and $T_2$ would coincide
only in the limit $q\to 1$.

In contrast, when the entropy obeys an \emph{additive} composition rule, as we have shown by replacing the statistical independence assumption by  the correlated microstate states, then, at fixed total energy $E_1+E_2=\mathrm{const}$, the condition $\delta S_q=0$ yields
\begin{equation}\label{E24a}
\left(\frac{\partial S_q(1)}{\partial E_1}\right)_{V_1,N_1}
=
\left(\frac{\partial S_q(2)}{\partial E_2}\right)_{V_2,N_2},
\end{equation}
and combining this result with the thermodynamic definition (\ref{E23}), 
the standard zeroth law is recovered for all values of the $q$-parameter:
\begin{equation}\label{E25}
T_1=T_2.
\end{equation}
Therefore, when two q-gaseous subsystems prepared with the same $T$ are put in
thermal contact, no net energy flux occurs and the common $T$ is preserved; if their
initial temperatures differ, energy exchange drives them to a common $T$, recovering
the standard equilibrium condition $T_1=T_2=T$ in this Neo-Planckian setting. Consistently, the power-laws velocity distributions in (\ref{E9}) quantify the physical (thermodynamic) temperature $T$ of an ideal gas under correlations, without need to introduce any $q$-dependent “effective temperature”. Thus, when the microstates of  $W$ are correlated through the deformed composition adopted in (\ref{E10}), an entropic additivity of $S_q$ is obtained, thereby preserving the zeroth law of thermodynamics.

\section{Most probable distributions: The reverse case for Power-laws}
The equilibrium power laws in Eq.~(\ref{E9}) were obtained from a
\textit{bottom-up} Neo-Maxwellian approach, without invoking an entropy
functional. Reciprocally and also for consistency, the same velocity distribution function must also be
derived from a \textit{``top-down"} route, based on the method of the most
probable distribution. This method goes back to Boltzmann's probabilistic
description of a gas \cite{Boltzmann1877}. In this view, a macroscopic state is
specified by the occupation of elementary states available to the particles.
Such a macroscopic occupation pattern can be realized by many microscopic arrangements \(W\). The most probable distribution
is then the one associated with the largest \(W\).

For a dilute monatomic gas, where the energy cells become increasingly narrow
and densely distributed, the discrete occupation probabilities are 
replaced by a smooth velocity distribution, \(f_q(v)\), and the microscopic
energy associated with each velocity state is \(\varepsilon=mv^2/2\). Thus, in
this continuous representation, the \(q\)-entropy functional, understood as
the continuum counterpart of the generalized Boltzmann principle
\(S_q=k_B\ln_q W\), is written as
\begin{equation}
S_q = -k_B\int f_q(v)\,\ln_q f_q(v)\,d^3v .
\end{equation}
As usually, the most probable power-law distribution is obtained by the extreme of this functional
under particle number density and energy constraints:
\begin{equation}
    \int f_q(v)\, d^3v = n,
    \qquad
    \int  f_q(v) \varepsilon(v)\, d^3 v = U,
\end{equation}
where by definition $\varepsilon = {m v^2}/{2}$. 
Following standard lines, we consider the constrained
functional with two Lagrange multipliers.
Without altering the physical constraints, these
multipliers can be conveniently parametrized as
$\lambda\equiv (2-q)\alpha_q-1$ and
$\gamma\equiv (2-q)\beta_q$.
Thus, we write
\begin{align}
\label{eq:constrained_functional}
G_q[f]
&=\frac{S_q[f]}{k_B}
-\lambda\int f_q\,d^3v
-\gamma\int f_q\varepsilon(v)\,d^3v.
\end{align}
Now, using the variational identity and expressing the result in terms of $\alpha_q$ and $\beta_q$, we obtain
\begin{equation}
\delta\big[f_q(v)\ln_q f_q(v)\big]
=
\big[1+(2-q)\ln_q f_q(v)\big]\delta f_q(v),
\end{equation}
the stationarity condition, after simplification, becomes
\begin{equation}
\left[
\ln_q f_q(v)
+\alpha_q
+\beta_q\varepsilon(v)
\right]\delta f_q(v)=0.
\end{equation}
Consequently,
\begin{equation}
\label{E32}
\ln_q f_q(v)
=
-\alpha_q
-\beta_q\frac{mv^2}{2}.
\end{equation}

Since \(\delta f_q(v)\) is arbitrary, the term inside brackets must vanish thereby determining the values of the pair ($\alpha_q, \beta_q$) and, naturally,  the searched $f_q(v)$ distribution function. However, without loss of generality, it is more interesting to obtain a remarkable simplified  version through  a  kind of a ``gauge" -- like reparametrization of the original constrained  problem. To begin with, we recall the known identity for the product rule  of two q-logarithmic functions, namely:
\begin{align}
&e_q(x)\,e_q(y) = e_q [x+y + (1-q)xy],\\
&\ln_q(xy)= \ln_q x + \ln_q y + (1-q)\ln_q x\,\ln_q y.
\end{align}
Motivated by these identities, we seek a canonical
factorized representation of the equilibrium distribution 

\begin{equation}
\label{E33}
f_q(v)
=
e_q(-\alpha)\,
e_q\left(-\beta\frac{mv^2}{2}\right),
\end{equation}
where $\alpha$ and $\beta$ are new parameters related
to the original Lagrange multipliers.
Applying the $q$-logarithm to
Eq.~(\ref{E33}) and using
the product rule, we obtain
\begin{align}
\ln_q f_q(v)
&=
-\alpha-\beta\frac{mv^2}{2}
+(1-q)\alpha\beta\frac{mv^2}{2}
\nonumber\\
&=
-\alpha
-\beta\left[1-(1-q)\alpha\right]\frac{mv^2}{2}.
\label{E37}
\end{align}
Comparing Eqs.~(\ref{E32})
and (\ref{E37}), the coefficients
independent and proportional to the kinetic
energy must coincide. Therefore,
\begin{equation}
\label{eq:gauge-relations}
\alpha=\alpha_q,
\qquad
\beta=\frac{\beta_q}{1-(1-q)\alpha_q}.
\end{equation}

The normalization factor \(e_q(-\alpha)\) is determined by enforcing
\(\int f_q(v)\,d^3v=n\). Using the usual identification of the energy scale,
\(\beta=(k_B T)^{-1}\), one obtains
\begin{equation}\label{Ea2}
e_q(-\alpha) \;=\;
\begin{cases}
\displaystyle
n\left(\frac{m(1-q)}{2\pi k_B T}\right)^{\!3/2}
\frac{\,\Gamma\!\left(\frac{1}{1-q}+\frac{5}{2}\right)}{\Gamma\!\left(\frac{1}{1-q}+1\right)},
\\[10pt]
\displaystyle
n\left(\frac{m(q-1)}{2\pi k_B T}\right)^{\!3/2}
\frac{\,\Gamma\!\left(\frac{1}{q-1}\right)}{\Gamma\!\left(\frac{1}{q-1}-\frac{3}{2}\right)},
\end{cases}
\end{equation}
Inserting (\ref{Ea2}) into (\ref{E33}) the two distributions in (\ref{E9}) are recovered.
Finally, if the present calculations are independently verified, we emphasize that the additivity of Tsallis entropy represents a new and nontrivial feature deserving detailed theoretical scrutiny. Indeed, beyond resolving the long-standing controversy surrounding the zeroth law of thermodynamics within Tsallis’ framework, the power-law \textit{``bottom-up"} formulation provides a coherent bridge between $q$-kinetic theory and $q$-statistical mechanics, while its closing connection to the  \textit{``top-down"} reverse approach (including correlations among the microstates and entropic additivity are assumed (see P1 and P2 - becomes  more firmly established.

\noindent\textbf{Acknowledgments:} JASL is partially supported by CNPq (309636/2026-4) and MHB by FAPESP/CNPq (24/14163-3).
\appendix 
\section{Extended Boltzmann Counting for $S_q$}

It is easy to show that
the Boltzmann number counting also provides the same results of section V.  Consider
$N$ particles distributed among $K$ accessible cells, with occupations
$\{n_i\}$ satisfying $\sum_i n_i=N$.  In the independent limit, the
multiplicity is $W_{\mathrm B}=N!/\prod_i n_i!$.  For correlated states, we
define its $q$-deformed counterpart by
\begin{equation}
\label{A1}
 W_q \equiv e_q\!\left(\ln_q N!_q-\sum_{i=1}^K\ln_q n_i!_q\right),
\end{equation}
where the factorial is built using the same deformed algebra as the
composition rule: $N!_q = e_q\left[
\ln_q(N) + \ln_q(N-1) + \cdots + \ln_q(1)
\right]$. These definitions reduce
to the ordinary factorial and multinomial multiplicity when $q\to1$.
The corresponding entropy can be written as
\begin{equation}
\label{A2}
 \frac{S_q}{k_B}=\ln_q W_q
 =\ln_q N!_q-\sum_{i=1}^K\ln_q n_i!_q.
\end{equation}
Now, for macroscopic occupations, we define the step-like integral
approximation to the deformed factorial 
\begin{align}
\label{A3}
 \ln_q N!_q
 \simeq \int_1^N\ln_q x\,dx =N\ln_q\!\left(\frac{N}{C}\right)
       -\ln_q\!\left(\frac{1}{C}\right),
\end{align}
where the constant is $C=(2-q)^{1/(1-q)}$ so that  $\lim_{q\to1}C \to e$.
For $q\to1$ it reduces to Boltzmann relation
$\ln N!\simeq N\ln(N/e)+1$.  Inserting Eq.~(\ref{A3})
into Eq.~(\ref{A2}) and neglecting the cell-independent
boundary term in the macroscopic approximation yields
\begin{equation}
\label{A4}
 \frac{S_q}{k_B}\simeq
 N\ln_q\!\left(\frac{N}{C}\right)
 -\sum_{i=1}^K n_i\ln_q\!\left(\frac{n_i}{C}\right).
\end{equation}
The most probable occupations follow by extremizing
Eq.~(\ref{A4}) with the ordinary constraints $\sum_i n_i=N$, and $\sum_i n_i\varepsilon_i=U$. The constant $C$ is essential, since \begin{align}
\label{A5}
 \frac{d}{dn_i}\left[n_i\ln_q\!\left(\frac{n_i}{C}\right)\right]
 &=\frac{(2-q)(n_i/C)^{1-q}-1}{1-q}\\
 &=\frac{n_i^{1-q}-1}{1-q}=\ln_q n_i.\nonumber
\end{align}
Introducing the auxiliary Lagrange multipliers
$\alpha_q$ and $\beta_q$, stationarity of
$S_q/k_B-\alpha_q\sum_i n_i-\beta_q\sum_i n_i\varepsilon_i$
yields
\begin{equation}
\label{A6}
\ln_q n_i=-\alpha_q-\beta_q\varepsilon_i.
\end{equation}
Following the same gauge-like procedure introduced
in the main text, we seek the canonical factorized form
\begin{equation}
\label{A7}
n_i=e_q(-\alpha)e_q(-\beta\varepsilon_i).
\end{equation}
Applying the $q$-logarithmic product rule to Eq.~(\ref{A7})
and comparing with Eq.~(\ref{A6}), we obtain
$\alpha=\alpha_q$, $\beta=\beta_q/[1-(1-q)\alpha_q]$.
Thus, the equilibrium occupations recover the ordinary
Boltzmann form,
$n_i=\exp(-\alpha)\exp(-\beta\varepsilon_i)$,
in the limit $q\to1$.

\end{document}